\documentclass{moriond}
\usepackage[utf8]{inputenc}
\usepackage[DIV=13]{typearea}
\usepackage{amsmath,amsfonts,bbm,mathrsfs,amssymb}
\usepackage{bbold}
\usepackage{slashed,cancel}
\usepackage[usenames,dvipsnames]{xcolor}
\usepackage{cite}
\usepackage{bm} 
\usepackage{hyperref}
\hypersetup{colorlinks=true,urlcolor=Magenta,anchorcolor=blue,citecolor=blue,filecolor=blue,
            linkcolor=Magenta,menucolor=blue, linktocpage=true,pdfproducer=medialab}
\usepackage[english]{babel}
\usepackage{catchfile}
\usepackage{indentfirst}
\usepackage{graphicx}
\usepackage{float}
\usepackage{enumerate}
\usepackage{caption}
\usepackage{subfigure}
\usepackage{multirow}
\usepackage{booktabs}
\usepackage{physics}
\usepackage{nicefrac}
\usepackage{tikz-feynman}
\usepackage{pdflscape}
\usepackage{wrapfig,lipsum,booktabs}

\def\be{\begin{equation}}
\def\ee{\end{equation}}
\def\bea{\begin{eqnarray}}
\def\eea{\end{eqnarray}}

\def\tH{\widetilde{H}}

\newcommand{\cM}{\mathcal{M}}

\newcommand{\sL}{\mathscr{L}}
\newcommand{\hc}{\text{h.c.}}
\newcommand{\ov}[1]{\overline{#1}}

\newcommand{\eV}{\ \text{eV}}

\newcommand{\GeV}{\ \text{GeV}}

\newcommand{\Photo}{\includegraphics[height=35mm]{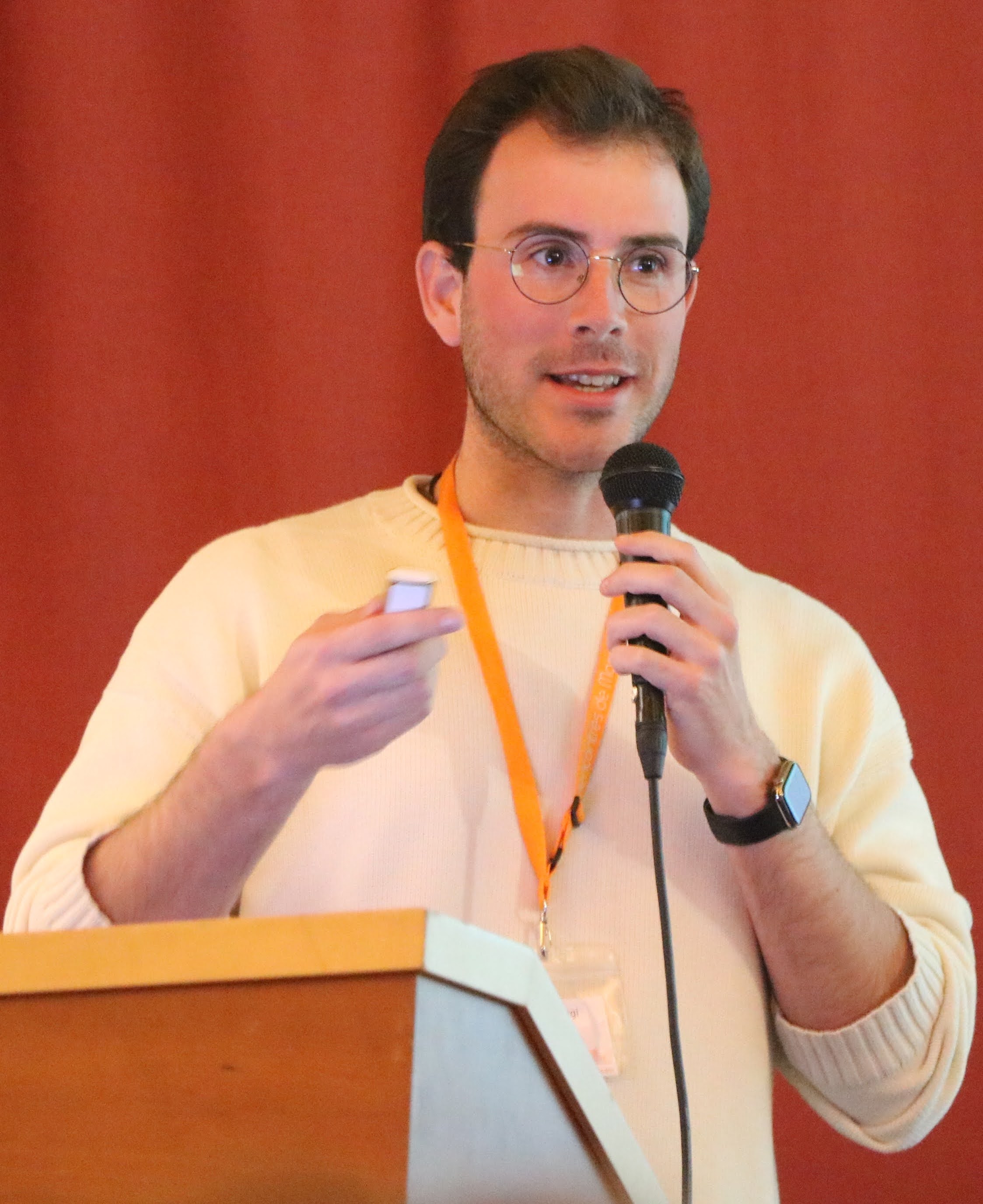}}

\begin{document}
\vspace*{4cm}
\title{Low Seesaw Scale Solution for $M_W$ and $(g-2)_\mu$}

\author{A. DE GIORGI$^{a,\ast}$, L. MERLO$^{a}$, S. POKORSKI$^{b}$}

\address{
${}^{a)}$ Departamento de F\'isica Te\'orica and Instituto de F\'isica Te\'orica UAM/CSIC,\\
Universidad Aut\'onoma de Madrid, Cantoblanco, 28049, Madrid, Spain
\vskip .2cm
${}^{b)}$ Institute of Theoretical Physics, Faculty of Physics,\\ 
University of Warsaw, Pasteura 5, PL 02-093, Warsaw, Poland
}

\maketitle\abstracts{
In this short talk, we present a renormalizable model that can i) generate neutrino masses via a low-scale seesaw mechanism and ii) solve the long-standing $(g-2)_\mu$ and the more recent CDF II $M_W$-anomalies. This is minimally achieved by introducing two sterile neutrinos and a single electroweak-doublet vector-like lepton, with masses $< 2$ TeV. We focus on the one-generation scenario and the requirements to extend it to three generations.}

\section{Introduction}
\label{sec:intro}
\newlength{\oldintextsep}
\setlength{\oldintextsep}{\intextsep}
\setlength\intextsep{5pt}
\begin{wraptable}[13]{r}{5.9cm}
\begin{tabular}{c|ccc} 
\toprule
&  $SU(2)_L $ & $U(1)_Y$ & $U(1)_L$\\
\midrule
$\ell_L$  	& $\bf 2$ 		& $-1/2$    & 1 \\
$\mu_R$   	& 1  			& 1	        & 1	\\
$H$       	& $\bf 2$ 		& $+1/2$ 	& 0 \\
\midrule
$N_R$  		& 1  		    & 1 		& 1  \\
$S_R$    	& 1  			& 1 		& -1 \\
$\psi_{L,R}$   	& $\bf2$  		& $-1/2$	& 1	\\
\bottomrule
\end{tabular}
\caption{\em Transformation properties under the gauge EW symmetry and their lepton charges.}
\end{wraptable}
Here we summarize the work presented in Ref.~\cite{deGiorgi:2022xhr}.

\noindent
Although we are aware of many problems and puzzles in the Standard Model, their resolution and understanding seems far off due to the lack of experimental data to guide us. 
However, it is fair to say that a single type of extension of the SM cannot solve them all if taken seriously. This prompts us to consider extensions of the SM that include at least two types of NP. If minimal or single-field extensions already left open an enormous number of possibilities, with two these are virtually infinite.
This prompts us to consider models where the new fields i) solve multiple problems or puzzles simultaneously and ii) have characteristics that stand out mutually and make the model more easily falsifiable.
Although anomalies are generically controversial and may, whether because they are the result of statistical fluctuations, experimental or theoretical error, lead to closed roads, they are very valuable to the scientific community. Both experimentally and theoretically, they prompt us to look for new ways to do research. They push us in constructing models to look for the most minimal extensions possible that work well together by exploiting the peculiar characteristics of each. This often results in flourishing areas, as in the case of Leptoquarks, studied in great detail for the $B$-Anomalies.

Motivated by this, we propose here a model that can explain, with the addition of only two vector-like leptons, neutrinos masses and two anomalies/tensions:
\begin{enumerate}
\item Fermilab's latest measurement of the anomalous magnetic moment of the muon, the $(g-2)_\mu$, confirmed a deviation from the SM prediction of
    $\delta a_\mu = (2.51\pm0.59)\times 10^{-9}$~\cite{Muong-2:2021ojo} ($4.2\sigma$);
\item The measurement of the mass of the W boson by the CDF II collaboration~\cite{CDF:2022hxs}
$M_W=80.4335\pm0.0094\GeV$,
which is in tension at $7\sigma$ with the SM prediction and with the previous ones from LHCb and ATLAS.
\end{enumerate}
\section{The Model}
\label{sec:model}
The SM is extended by two vector-like leptons: an EW doublet and a singlet under the SM gauge-group.
The Yukawa sector of the Lagrangian reads
\begin{equation}
\begin{aligned}
-\sL_Y=&\phantom{+}\ov{\ell_L}HY_\mu \mu_R+
\ov{\ell_L}\tH Y_N N_R+
\epsilon\ov{\ell_L}\tH Y_S S_R+
\dfrac{1}{2}\mu\ov{S_R^c}S_R+
\Lambda\ov{N_R^c}S_R+
+Y_R\ov{\psi_L}H\mu_R\\
&+
Y_V\ov{S_R^c}\tH^\dag\psi_R+
Y'_V\ov{\psi_L}\tH N_R
+
M_\psi\ov{\psi_L}\psi_R+
M_L\ov{\ell_L}\psi_R+
\hc\,,
\end{aligned}
\label{1GenLag}
\end{equation}
where $\ell_L$ and $\mu_R$ are the SM leptons, $H$ is the Higgs doublet, $N_R$ and $S_R$ are the two sterile neutrinos, $\psi$ is the vector-like EW doublet, and finally $\tH\equiv i\sigma_2H^\ast$, with $\sigma_2$ the second Pauli matrix. All the terms respect lepton number conservation, except for those proportional to $\epsilon Y_S$ and $\mu$.
We work in perturbation theory expanding the results in powers of $v/(\Lambda, M_\psi, M_L)$.
For consistency we consider only the case $\Lambda,\, M_{\psi,L}\gtrsim500\GeV$.
\section{Phenomenology}
\paragraph{Neutrino Masses}
The neutrino masses can be generated via a Low-Scale SS mechanism~\cite{Wyler:1982dd, Mohapatra:1986bd, Bernabeu:1987gr, Malinsky:2005bi, Abada:2007ux}. Once the EW symmetry is spontaneously broken, the mass terms can be rewritten as
\begin{equation}
-\sL_Y\supset
\dfrac{1}{2}\ov{\chi}\cM_\chi\chi^c+
\ov{\zeta_L}\cM_\zeta\zeta_R+
\hc\,,
\end{equation}
with $\chi\equiv(\nu_L,\, N_R^c,\, S_R^c,\,\psi^0_L,\,\psi_R^{0c})^T$, 
$\zeta\equiv(\mu,\,\psi^-)^T$ and
\begin{equation}
\cM_\chi= 
    \left(
    \begin{matrix}
        0   & m_N       & \epsilon\, m_S       & 0         & M_L \\
        m_N & 0         & \Lambda   & m_{V'}    & 0 \\
        \epsilon\, m_S & \Lambda   & \mu       & 0         & m_V \\
        0   & m_{V'}    & 0         & 0         & M_\psi  \\
        M_L & 0         & m_V       & M_\psi    & 0 \\
    \end{matrix}
    \right)\,,\qquad
\cM_\zeta= 
\left(
\begin{matrix}
m_\mu & M_L\\
m_R & M_\psi
\end{matrix}
\right)\,,
\end{equation}
where $m_i\equiv \dfrac{Y_{i}v}{\sqrt{2}}$. The parameter $M_L$ can be eliminated by a rotation of $\cos{\theta}\equiv \dfrac{M_\psi}{\widetilde{M}_\psi}$ with $
\widetilde{M}_\psi \equiv \sqrt{M_\psi^2+M_L^2}$. This amounts to a redefinition of the original parameters, $m_{\mu,R,N,V'}\to \widetilde{m}_{\mu,R,N,V'}$. We will denote the fields in the mass basis with $\widehat{\cdot}$.

In this simplified model, only one active neutrino gets mass. We arbitrarily choose that its mass corresponds to the square root of the atmospheric mass squared difference, $\widehat{m}_\nu\sim\sqrt{\Delta m^2_\text{atm}}$~\cite{Esteban:2020cvm}.
Given the range on $\widetilde{m}_N/\Lambda$ to explain the CDF II measurement of $M_W$, one needs $\mu-15\,\epsilon\,Y_s\,v\cos\theta\approx6\eV$.
\begin{figure}[tbh]
\centering
\subfigure[{\em Diagrams contributing to the $g-2$ of the muon at 1-loop in unitary gauge.}\label{fig:g-2-diagrams}]{
\includegraphics[width=0.43\textwidth]{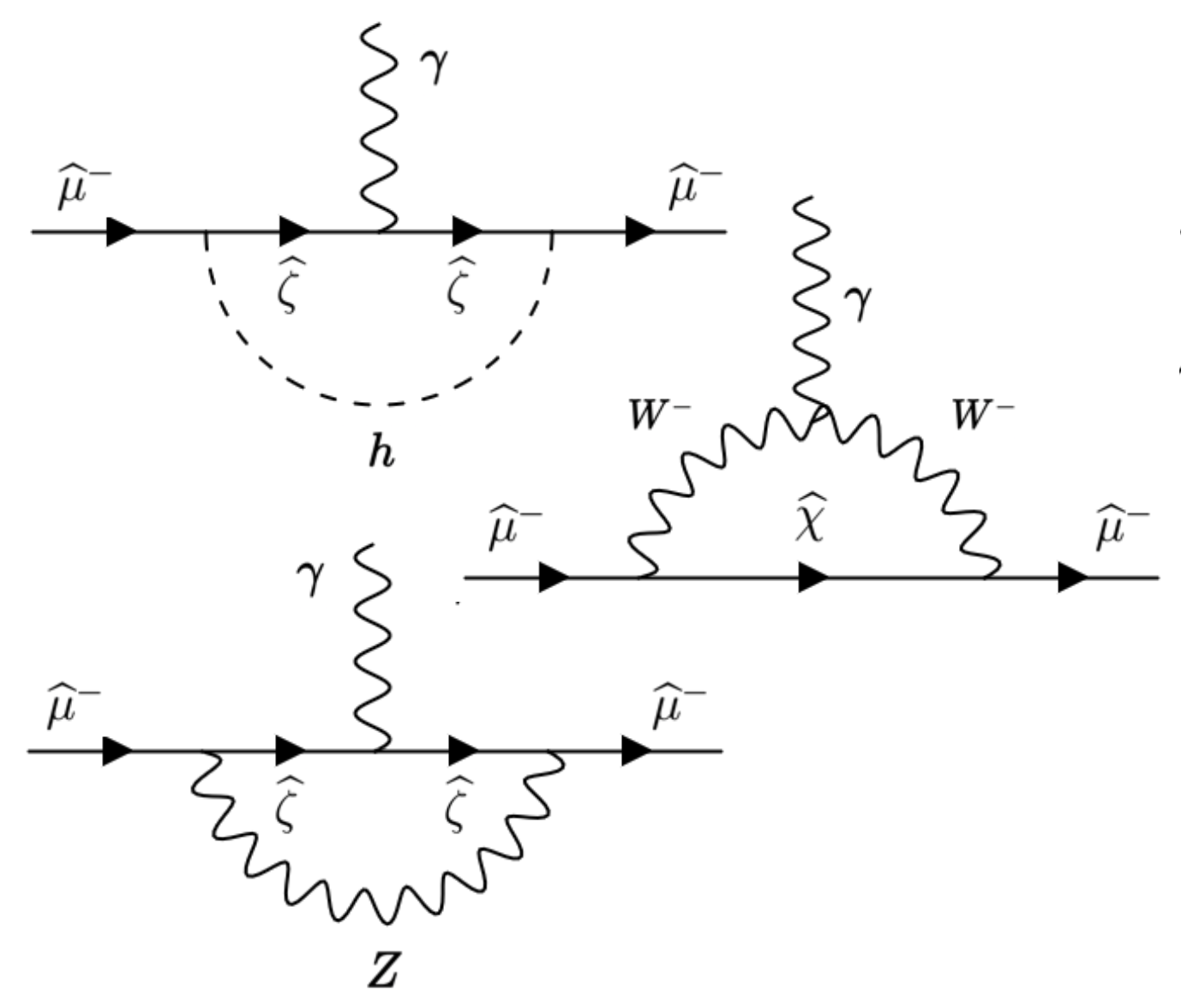}}
\quad
\subfigure[{\em The black curves represent the ratio $\delta m_\mu/m_\mu^{\exp}$. $\delta a_\mu$ is set to the experimental central value.}\label{fig:deltamuvsmuexp}]{
\includegraphics[width=0.43\textwidth]{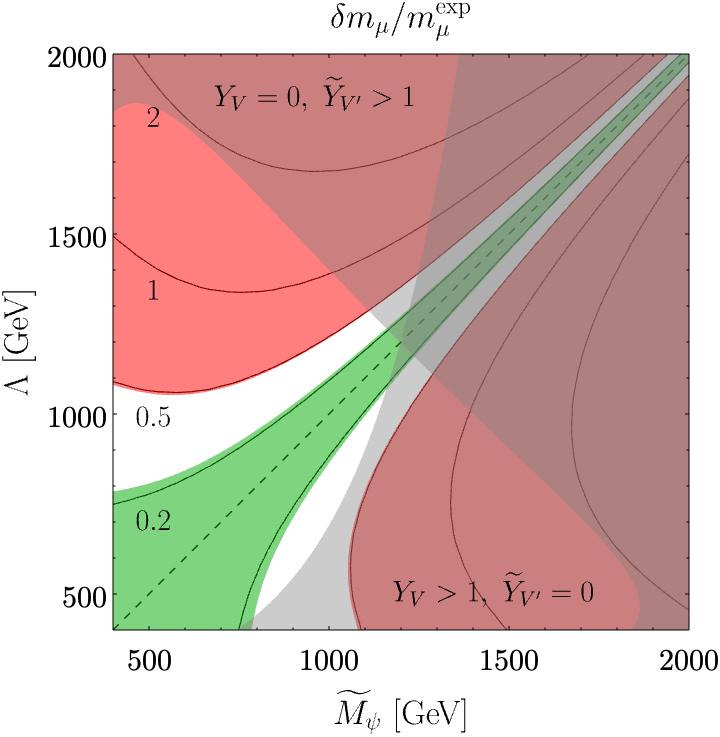}}
\caption{}
\end{figure}
\vspace{-10pt}
\paragraph{\boldmath CDF II $M_W$ Tension}
The modification of the $W-\mu$ coupling induced by the mixing with the sterile neutrinos has an impact on the computation of the  muon $\beta$ decay
\begin{equation}
    \Gamma_\mu \simeq \dfrac{m^{\text{exp}\,5}_\mu G_F^2}{192\,\pi^3}\left(1-\dfrac{\widetilde m_N^2}{2\Lambda^2}\right)\equiv\dfrac{m^{\text{exp}\,5}_\mu G_\mu^2}{192\,\pi^3}\,,\qquad     G_F 
    \simeq G_\mu \left(1+\dfrac{\widetilde m_N^2}{2\Lambda^2}\right) \,,
\end{equation}
where $G_F$ is the Fermi constant parameter as defined in the Fermi Lagrangian, and $G_\mu$ is the experimental determination extracted from the muon lifetime. 
This implies a modification of the relation between the $W$ boson mass and the experimental determination of $G_\mu$~\cite{Blennow:2022yfm, Arias-Aragon:2022ats},
\begin{equation}
    M_W\simeq M_Z\,
    \sqrt{
    \dfrac{1}{2}+
    \sqrt{
    \dfrac{1}{4}-
    \dfrac{\pi\,\alpha_\text{em}}{\sqrt2\,G_\mu\,M_Z^2\,\left(1-\Delta r\right)}
    \left(1-\dfrac{\widetilde m_N^2}{2\Lambda^2}\right)
    }
    }\,.
    \label{MWPrediction1Gen}
\end{equation}
The CDF II measurement is explained for 
$
\dfrac{\widetilde m_N^2}{\Lambda^2}\in [6.6,11.8]\cross 10^{-3} \, \text{at}~2\sigma$.
Loop-level modifications are subleading with respect to the tree-level one.
\paragraph{Muon magnetic dipole moment, \boldmath $(g-2)_\mu$}
The NP contributions  to  $(g-2)_\mu$,
are associated with the Feynman diagrams in Fig.~\ref{fig:g-2-diagrams}. The same diagrams without the external photon generate the 1-loop correction to the muon mass, $\delta m_\mu$.
At $1$-loop, the chirally enhanced LO, i.e. $\mathcal{O}(v^2/(\Lambda,\widetilde{M}_\psi)^2)$, contribution to $\delta a_\mu$ vanishes~\cite{Arkani-Hamed:2021xlp, Craig:2021ksw}.
 Defined $F_0(x,y)\equiv \dfrac{3}{2}-\dfrac{x\log{y}-y\log{x}}{x-y}$, the NLO is suppressed by four powers of the heavy neutral masses and reads
\begin{equation}
  \delta a_\mu^{\text{CE-1L}}=
    \dfrac{3\,m^\text{exp}_\mu}{4\,\pi^2\,v^2}
    \dfrac{M_W^2}{\Lambda\widetilde{M}_\psi}
    \dfrac{\widetilde{m}_N\widetilde{m}_R}{\widetilde{M}_\psi}
    \left(\dfrac{m_V}{\widetilde{M}_\psi}+\dfrac{\widetilde{m}_{V'}}{\Lambda}\right)
    \,F_0\left(\dfrac{\Lambda^2}{M_W^2},\dfrac{\widetilde{M}_\psi^2}{M_W^2}\right)\,.
\label{g2muCE1LNLO}    
\end{equation}
The $2$-loop diagrams turn out to be smaller than the $20\%$ of the LO contribution in the considered parameter space. 

\section{Conclusions}
The results are shown in Fig.~\ref{fig:deltamuvsmuexp}. In the green(white)[red] region $\delta m_{\mu}<0.3$($0.3<\delta m_{\mu}< 1$)[$\delta m_{\mu}>1$] the tree-level mass. In grey are shown the regions where $1<|Y_V|<5$ with $\widetilde{Y}_{V'}=0$, or  $1<|\widetilde{Y}_{V'}|<5$ with $Y_V=0$.
The correct value of $\delta a_\mu$ and of $M_W$ is achieved with vectorlike leptons of masses $\lesssim 2$ TeV without the necessity of an unnaturally large cancellation between tree- and loop-level contributions to the muon mass. This allows for a very rich phenomenology at the TeV-scale at collider.
The one-generation scenario studied here is severely constrained but can be eased in the more realistic three-generations model. In such a case, further flavour structure must be included in the model to protect it from the strong bounds from $\mu\to e \gamma$.

\section*{Acknowledgments}
A.d.G. and L.M. acknowledge partial financial support 
through the grant IFT Centro de Excelencia Severo Ochoa No CEX2020-001007-S by the grant PID2019-108892RB-I00 funded by MCIN/AEI/ 10.13039/501100011033, 
and the grant agreement No 860881-HIDDeN. 
The research of S.P. has received partial financial support 
through the grant DEC-2016/23/G/ST2/04301.

\end{document}